\documentclass[letterpaper]{article} 
\usepackage{aaai25}  
\usepackage{times}  
\usepackage{helvet}  
\usepackage{courier}  
\usepackage[hyphens]{url}  
\usepackage{graphicx} 
\usepackage{natbib}  
\usepackage{caption} 
\usepackage{algorithm}
\usepackage{algorithmic}
\usepackage{booktabs}
\usepackage{subcaption}
\usepackage{amsmath}
\usepackage{amssymb}
\usepackage{newfloat}
\usepackage{listings}
\DeclareCaptionStyle{ruled}{labelfont=normalfont,labelsep=colon,strut=off} 
\floatstyle{ruled}
\newfloat{listing}{tb}{lst}{}
\floatname{listing}{Listing}
\title{Will AI Take My Job?
Evolving Perceptions of Automation and Labor Risk in Latin America}
\author {
    Andrea Cremaschi\textsuperscript{\rm 1},
    Dae-Jin Lee\textsuperscript{\rm 1},
    Manuele Leonelli\textsuperscript{\rm 1}
}
\affiliations {
    \textsuperscript{\rm 1}School of Science and Technology, IE University, Madrid, Spain\\
    andrea.cremaschi@ie.edu, dae-jin.lee@ie.edu, manuele.leonelli@ie.edu
}

\usepackage{bibentry}

\begin{document}

\maketitle

\begin{abstract}
As artificial intelligence and robotics increasingly reshape the global labor market, understanding public perceptions of these technologies becomes critical. We examine how these perceptions have evolved across Latin America, using survey data from the 2017, 2018, 2020, and 2023 waves of the Latinobarómetro. Drawing on responses from over 48,000 individuals across 16 countries, we analyze fear of job loss due to artificial intelligence and robotics. Using statistical modeling and latent class analysis, we identify key structural and ideological predictors of concern, with education level and political orientation emerging as the most consistent drivers. Our findings reveal substantial temporal and cross-country variation, with a notable peak in fear during 2018 and distinct attitudinal profiles emerging from latent segmentation. These results offer new insights into the social and structural dimensions of AI anxiety in emerging economies and contribute to a broader understanding of public attitudes toward automation beyond the Global North.
\end{abstract}

%
 \begin{links}
     \link{Extended version}{https://arxiv.org/abs/2505.08841}
 \end{links}

\section{Introduction}

As artificial intelligence (AI) technologies become increasingly embedded in the fabric of economic and social life, public concern about their long-term implications has grown. Among the most salient fears is automation displacing human labor, particularly in economies marked by informality, inequality, and institutional fragility. While techno-optimist narratives highlight gains in productivity and efficiency, they often overlook the uneven social consequences of AI deployment, especially fears about job loss, economic marginalization, and weakened democratic accountability.

Existing research has shown that public attitudes toward AI are shaped by a complex interplay of individual characteristics, political orientations, and broader cultural and institutional contexts \citep{bergdahl2023selfdetermination, gillespie2021trust, schiavo2024comprehension}. Several studies have found that education, age, gender, and political ideology influence perceptions of AI \citep{gerlich2023perceptions, zhang2019ai, kaya2024personality}, while others emphasize the role of institutional trust in shaping public acceptance \citep{gillespie2021trust, schiavo2024comprehension}. However, most of this evidence is concentrated in North America, Europe, or East Asia. Relatively little is known about how populations in Latin America, where economic precarity and political polarization are prevalent, perceive the labor risks associated with AI.

This paper addresses this gap by analyzing responses from over 48,000 individuals across 16 Latin American countries, drawn from the four most recent waves (2017, 2018, 2020, and 2023) of the Latinobarómetro survey \citep{latinobarometro}. Using a combination of statistical modeling and latent class analysis, we examine how fear of AI-induced job loss varies across countries, time periods, and social groups. Our analysis highlights education level and political ideology as the most consistent predictors of concern, while also uncovering how these attitudes cluster into distinct societal profiles.

We find that concern about AI-driven job loss peaked in 2018, likely due to changes in question framing and heightened media attention, but remains widespread and unequally distributed. Fear is especially pronounced among individuals with lower educational attainment. Political ideology also plays a role, with left-leaning respondents expressing greater concern in earlier years, though this pattern becomes less consistent over time and, in some cases, reverses. The results underscore the importance of structural conditions and ideological identity in shaping public perceptions of automation in Latin America. In doing so, this study contributes a comparative perspective to a literature still heavily focused on the Global North, offering new insight into how technological anxiety is patterned in emerging economies.

\section{Literature Review}

Public concern about AI displacing human workers is a dominant theme in public discourse, policy debates, and empirical studies. However, how this fear varies across countries and social groups, and how it relates to institutional trust and economic perceptions, remains an active area of research. In this section, we review the literature across five main themes relevant to our study: general fears of job loss due to AI, geographic differences, sociodemographic variation, institutional trust, and economic outlook.

\subsection*{Fear of Job Loss to AI}

Fears about AI-induced job loss are widespread. \citet{frey2017future} estimated that up to 47\% of U.S. jobs could be at risk of automation, while \citet{acemoglu2017robots} estimated annual job losses of 360,000--670,000 due to robotics. Recent work confirms these concerns are now deeply embedded in public consciousness.  \citet{gerlich2023perceptions} and \citet{kaya2024personality} report that the threat of job displacement ranks among the most salient AI-related fears \citep{chiarini2023ai,li2020dimensions,nnamdi2023impact,soueidan2024impact,wilde2025fear}. In the 2024 UK Tracker Survey, over 70\% of respondents voiced concern about AI-induced unemployment \citep{dsit2024wave4}. However, \citet{schiavo2024comprehension} show that job replacement anxiety, while common, is not a strong predictor of AI acceptance \citep{ucar2024relationship}.

\subsection*{Geographic and Cultural Variation}

The intensity of AI-related fears varies across regions. European respondents tend to be more skeptical of AI than their counterparts in emerging economies \citep{gillespie2021trust}. \citet{bergdahl2023selfdetermination} report that AI positivity is highest in Finland and Poland and lowest in France, suggesting clear cultural and institutional differences. Latin America remains understudied, but evidence from the Latinobarómetro surveys reveals a distinct pattern: concern about AI-driven job loss peaked sharply in 2018 before stabilizing, as we demonstrate in our longitudinal analysis below. This trajectory mirrors findings from the UK and Germany, where concern tends to rise in response to media and policy events but attenuates with increased familiarity and exposure \citep{dong2024fears,gerlich2023perceptions}.

\subsection*{Sociodemographic Predictors of AI Fear}

\textbf{Age.} Studies find that younger individuals hold more positive attitudes toward AI \citep{zhang2019ai, bergdahl2023selfdetermination}, yet this effect may be mediated by digital literacy and perceived self-efficacy rather than age itself \citep{schiavo2024comprehension}. However, some studies suggest younger individuals may also perceive greater long-term vulnerability to automation \citep{morikawa2017afraid}.

\textbf{Gender.} Women consistently express greater anxiety about AI, particularly regarding fairness, job loss, and ethical misuse \citep{bergdahl2023selfdetermination,kaya2024personality}. \citet{kaya2024personality} show that women report significantly higher levels of AI learning and job replacement anxiety.

\textbf{Education.} Education is a robust predictor of AI attitudes. Higher education is associated with lower anxiety and greater acceptance \citep{schiavo2024comprehension}. AI literacy, defined as understanding, using, and critically evaluating AI, is both shaped by and mediates the effect of education on AI acceptance \citep{lund2024ai}.

\textbf{Ideology.} Political ideology shapes concerns about AI. Right-leaning individuals tend to focus on cultural threats and loss of control, whereas left-leaning respondents emphasize labor rights and inequality \citep{gerlich2023perceptions}. These findings echo broader ideological divides in technology governance \citep{bergdahl2023selfdetermination}.

\textbf{City Size and Technological Exposure.} Existing research suggests that residents of larger cities are generally more exposed to AI and tend to hold more nuanced views, often combining optimism with concern \citep{dsit2024wave4}. In contrast, individuals living in smaller towns or rural areas frequently express greater skepticism, likely due to lower digital engagement and limited perceived benefits.

\subsection*{Trust in Institutions and Governance}

Trust in those developing and governing AI is a strong predictor of public acceptance. While trust in academia and healthcare is relatively high, governments and tech companies are viewed more skeptically \citep{gillespie2021trust}. This trust gap fuels anxiety: respondents want regulation but question government competence \citep{cremaschi2025understanding,gerlich2023perceptions,montag2023propensity}, and anxiety drops when governance is trusted \citep{schiavo2024comprehension,zhan2024what}. A lack of trust heightens concerns about autonomy, transparency, and job loss.

\subsection*{Economic Outlook and Risk Perceptions}

Economic perceptions, both macro and personal, play a significant role in AI fear. Individuals who believe the economy is deteriorating are more likely to see AI as a threat to job security \citep{kaya2024personality}. \citet{bergdahl2023selfdetermination} report that negative income expectations are strongly correlated with AI negativity. However, economic optimism can mitigate AI fear. For example, individuals who see AI as economically beneficial are more likely to accept it even when aware of its risks \citep{gerlich2023perceptions,george2024future}. Thus, economic context mediates whether automation is viewed as a threat or a tool \citep{gu2025fear}.

\medskip
In summary, the literature shows that fear of AI-driven job loss is widespread but shaped by demographic, cultural, and institutional factors. Education and trust, especially in government, emerge as key buffers against AI anxiety, while negative economic expectations heighten it. These insights frame our investigation of job loss fears in Latin America, where structural inequality and institutional fragility may shape attitudes in distinct ways.

\begin{table*}
\small 
\centering
\begin{tabular}{lll}
\toprule
\textbf{Variable} & \textbf{Description} & \textbf{Levels} \\
\midrule
\textbf{fear\_ai} & Fear of losing one's job due to AI/robots &
\textit{Strongly disagree, Disagree, Agree, Strongly agree} \\
\textbf{country} & Country of residence &
\textit{Bolivia, Brazil, ..., Venezuela} \\
\textbf{age} & Age of respondent & Numeric (continuous) \\
\textbf{gender} & Gender & \textit{Female, Male} \\
\textbf{education} & Education level & \textit{Low, Medium, High} \\
\textbf{city\_size} & Size of locality &
\textit{Small town, Mid-sized city, Large city, Capital} \\
\textbf{econ\_now} & Perception of current economy &
\textit{Bad, Average, Good} \\
\textbf{econ\_future} & National economic outlook &
\textit{Worse, Same, Better} \\
\textbf{econ\_personal\_future} & Personal/family economic outlook &
\textit{Worse, Same, Better} \\
\textbf{trust\_people} & Generalized trust &
\textit{Cannot be too careful, Trust most} \\
\textbf{democracy\_pref} & Support for democracy &
\textit{Doesn't matter, Authoritarian OK, Democracy best} \\
\textbf{demo\_satisfaction} & Satisfaction with democracy &
\textit{Not at all satisfied, Not very satisfied, Satisfied} \\
\textbf{trust\_parliament} & Trust in parliament &
\textit{None, Little, Some or A lot} \\
\textbf{trust\_gov} & Trust in government &
\textit{None, Little, Some or A lot} \\
\textbf{trust\_judiciary} & Trust in judiciary &
\textit{None, Little, Some or A lot} \\
\textbf{ideology\_group} & Political ideology &
\textit{Left, Centre, Right} \\
\textbf{year} & Year of response & \textit{2017, 2018, 2020, 2023} \\
\bottomrule
\end{tabular}
\caption{Variables used in the analysis, with descriptions and categorical levels after final cleaning.}
\label{tab:variables}
\end{table*}

\section{Materials and Methods}

\subsection{Dataset Description}
The analysis is based on survey data collected by \textit{Latinobarómetro}, a public opinion study conducted across Latin America. We use the four most recent editions of the survey (2017, 2018, 2020, and 2023), which include responses from 16 countries: Bolivia, Brazil, Chile, Colombia, Costa Rica, Dominican Republic, Ecuador, El Salvador, Guatemala, Honduras, Mexico, Panama, Paraguay, Peru, Uruguay, and Venezuela. Each wave captures perceptions of democracy, governance, economic outlook, and, crucially, emerging technologies like AI.

We analyze a subset of variables that are harmonized across waves and relevant to perceptions of AI. The outcome of interest is fear of job loss due to AI or robots, measured on a four-point Likert scale ranging from “Strongly disagree” to “Strongly agree.” Key predictors include demographic characteristics (age, gender, education, and city size), political orientation (left–centre–right self-placement), economic perceptions (views of the national economy, future national outlook, and future personal situation), and institutional attitudes (trust in people, government, parliament, and judiciary, as well as preferences for democracy and satisfaction with its functioning), with country and year included to capture geographic and temporal variation. Table~\ref{tab:variables} summarizes the chosen variables.

The question used to construct \textbf{fear\_ai} varies slightly across waves. In 2017, respondents were asked whether artificial intelligence and robots would reduce more jobs than they create. In 2018, they were asked whether robots would take their job or a family member’s job. In 2020, the question was whether robots would take away their workplace within 10 years, and in 2023, whether robots would take their job in 10 years’ time. While the exact wording differs across waves, all questions tap into a shared concern about the future impact of robotics and AI on employment. We treat these questions as conceptually equivalent for the purpose of harmonized modeling, while acknowledging the slight variation in framing.

\subsection{Data Preparation}

We apply a standardized preprocessing pipeline to ensure comparability across survey waves, including harmonizing variable names and formats, recoding categorical responses into consistent factor levels, merging sparse or semantically similar categories, and enforcing ordinal structures where applicable. All variables except age are encoded as factors using sum-to-zero contrast coding, and we conduct a complete-case analysis to retain only fully observed records. Two countries are excluded: Nicaragua, not included in the 2023 wave, and Argentina, due to systematic missingness in key variables such as education and ideology in 2020. The final dataset comprises 48,403 individuals.

\subsection{Exploratory Analysis}

We conduct all statistical analyses using \texttt{R} within the \texttt{RStudio} environment \citep{rstudio}. We first perform exploratory data analysis to assess the distribution and bivariate associations of the response variable \textbf{fear\_ai} with the available predictors. We use the \texttt{ggplot2} package \citep{ggplot2} to produce stacked barplots displaying the distribution of \textbf{fear\_ai} across countries and years. Visualizations are faceted by country to aid comparison.

To quantify the strength of association between \textbf{fear\_ai} and each feature, we compute mutual information (MI) using the \texttt{infotheo} package \citep{infotheo}. To evaluate statistical dependence, we conduct Fisher’s exact tests of independence with simulated p-values. For the continuous predictor \textbf{age}, we discretize it into quartiles.

\subsection{Statistical Modeling}

To model the relationship between the outcome variable \textbf{fear\_ai} and the available predictors, we use \textit{ordinal logistic regression}, also known as the \textit{proportional odds model} \citep{agresti2013categorical,mccullagh1980regression}. This approach is well suited for outcomes measured on a four-point Likert scale.

In ordinal logistic regression, the model estimates the log-odds of being at or below a certain category, given a set of predictor variables. Formally, let $Y_i$ denote the ordinal response for individual $i$, taking values in $\{1, 2, ..., K\}$, with $K=4$  here. The model assumes:
\[
\log\left(\frac{\mathbb{P}(Y_i \leq k)}{\mathbb{P}(Y_i > k)}\right) = \theta_k - \mathbf{x}_i^\top \boldsymbol{\beta}, \quad \mbox{for } k = 1, 2, ..., K-1,
\]
where $\theta_k$ are $K-1$ threshold parameters separating the response categories, $\mathbf{x}_i$ is the vector of predictors for individual $i$, and $\boldsymbol{\beta}$ is the vector of regression coefficients. The model assumes that $\boldsymbol{\beta}$ is constant across the $K-1$ equations: this is known as the \textit{proportional odds assumption}. We implement this model using the \texttt{polr()} function from the \texttt{MASS} package in R \citep{MASS}.

All categorical predictors are encoded using \textit{sum-to-zero} constraints. This coding ensures that for any categorical variable with $J$ levels, the model includes $J$ coefficients whose sum equals zero. Each level's coefficient represents a deviation from the average effect of that variable, making comparisons across levels and countries more interpretable. Coefficient signs indicate direction: positive values imply greater likelihood of stronger fear responses, while negative values suggest the opposite. The magnitude of the coefficient reflects the strength of this relationship.

We fit separate models for each country and each year (2017, 2018, 2020, 2023), as well as a pooled ``FULL'' model per year that includes all countries. In each case, we use a \textit{greedy forward selection} algorithm to identify a parsimonious set of predictors. Starting from an intercept-only model, we iteratively add the variable that yields the largest improvement in out-of-sample log-likelihood, measured using 10-fold cross-validation. This procedure continues until no remaining variable improves model fit.

To quantify the strength of association between a predictor and the outcome, we use the \textit{average absolute coefficient} across all levels of the predictor.
This measure provides an intuitive summary of a variable's impact: the larger the average absolute coefficient, the greater the predictor's contribution to shifting individuals across fear categories.

This framework allows us to examine both whether a variable is selected for inclusion in a given country-year model, and how strongly it influences attitudes toward AI-induced job loss. 

\subsection{Latent Class Analysis for Profile Identification}
To uncover distinct latent profiles based on respondents’ characteristics and attitudes, we perform Latent Class Analysis (LCA). LCA is a probabilistic clustering method that identifies subgroups of individuals who share similar response patterns across categorical variables, under the assumption of conditional independence within classes \citep{lazarsfeld1950latent}. We include all predictors used in the study, except for \textbf{country} and \textbf{fear\_ai}, so that the latent classes reflect general patterns of demographic background, economic perceptions, political attitudes, and institutional trust, rather than being directly driven by geography or fear of AI-related job loss. The continuous variable \textbf{age} is again divided into quartiles.

LCA models the joint distribution of observed categorical variables \( \mathbf{X} = (X_1, X_2, \ldots, X_J) \) through a categorical latent variable \( C \in \{1, \ldots, K\} \), where \( K \) is the number of latent classes. Under the assumption of conditional independence, the model is specified as:
\[
\mathbb{P}(\mathbf{X} = \mathbf{x}) = \sum_{k=1}^K \pi_k \prod_{j=1}^J \mathbb{P}(X_j = x_j \mid C = k),
\]
where \( \pi_k = \mathbb{P}(C = k) \) is the proportion of individuals in class \( k \), and \( \mathbb{P}(X_j = x_j \mid C = k) \) is the item-response probability: the probability of a particular response on variable \( X_j \), conditional on membership in class \( k \).

These probabilities are estimated via maximum likelihood using the \texttt{poLCA} package in \texttt{R} \citep{linzer2011polca}. For each number of latent classes from 1 to 10, we fit models using 20 random starts to ensure robust convergence. For each model, we compute the Bayesian Information Criterion (BIC), which balances model fit and complexity, and the entropy, which quantifies classification certainty, with higher entropy indicating clearer separation between classes. We use both metrics to select a number of classes that fits the data well while producing distinct and interpretable profiles.

After selecting the optimal model, respondents are assigned to the class with the highest posterior membership probability. The estimated item-response probabilities are then used to characterize the substantive meaning of each latent profile. These classes form the basis for analyzing variation in AI-related job loss fear across societal segments. To assess whether \textbf{fear\_ai} distributions differ significantly between classes, we conduct chi-squared tests for each survey year, followed by pairwise comparisons with adjustment for multiple testing \citep{benjamini1995fdr}.

\section{Results}

\subsection{Overview of Fear of AI-Driven Job Loss}
Across the four waves of the Latinobarómetro survey, respondents expressed varying degrees of concern about the risk of job loss due to AI. Overall, the most common response was \textit{Agree} (45.6\%), followed by \textit{Disagree} (29.6\%). A smaller proportion of respondents reported \textit{Strongly agree} (17.5\%), while only 7.3\% selected \textit{Strongly disagree}, suggesting that fears related to AI are widespread but often moderate in intensity.

\begin{figure*}
    \begin{center}
\includegraphics{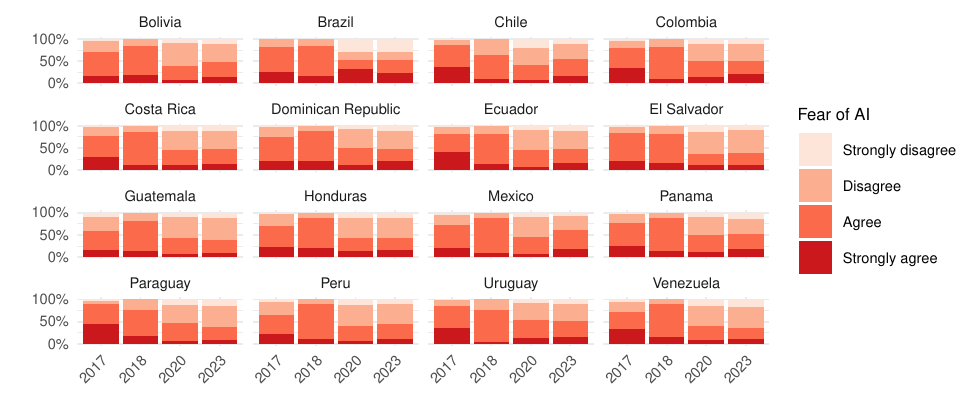}
    \end{center}
    \caption{Fear of AI-driven job loss across years, by country. Bars show the distribution of responses to \textbf{fear\_ai} from 2017 to 2023.}
    \label{fig:fearai_country}
\end{figure*}

When examining responses by year, there is notable variation. In 2018, concern was especially pronounced: nearly 69.3\% of respondents \textit{Agreed} that robots could take their job or a family member's, while only 14.3\% \textit{Strongly agreed} and no respondents selected \textit{Strongly disagree}. By 2020 and 2023, a more balanced spread across categories is observed, with both waves showing over 12\% of respondents selecting \textit{Strongly disagree}, and moderate concern (\textit{Disagree} or \textit{Agree}) distributed more evenly. Interestingly, in 2017, there was a greater proportion of respondents who expressed \textit{Strongly agree} (28.0\%) compared to any other year, pointing to an earlier polarization in views around AI-driven job displacement.

The evolution of concern about AI-driven job loss shows clear cross-country differences and dynamic patterns over time, as illustrated in Figure~\ref{fig:fearai_country}. Brazil stands out for its high polarization: across all waves, it exhibits a notably large share of both \textit{Strongly agree} and \textit{Strongly disagree} responses, much more so than any other country. This points to entrenched divisions over automation in Brazil’s labor market. In contrast, countries such as Venezuela and Paraguay show initially high levels of fear, particularly in 2017 and 2018, followed by a noticeable decline in 2020. This pattern may reflect shifting perceptions as automation discourse matured or became less salient during that period. Most other countries (with the notable exception of Brazil) display a U-shaped trend: relatively high concern in 2018, a dip in 2020, and a modest rebound in 2023. This dip may be partially attributed to a change in question framing. In 2018, respondents were asked whether robots could take their job \textit{or a family member’s job}, potentially inflating perceived risk, whereas in 2020 and 2023, the question focused exclusively on personal job loss. Nonetheless, the renewed rise in concern by 2023 suggests that fear of automation is resurging, likely fueled by increasing real-world adoption of AI technologies and heightened media attention.

Overall, while temporal patterns are broadly consistent across countries, the magnitude of fear and the degree of polarization vary substantially, underscoring the importance of contextual factors in shaping public perceptions of AI-driven labor market disruptions.

\begin{table*}
\small
\centering
\begin{tabular}{lrrrrr}
\toprule
\textbf{Feature} & \textbf{Overall} & \textbf{2017} & \textbf{2018} & \textbf{2020} & \textbf{2023} \\
\midrule
\textbf{country}                & 0.0112 (0.000) & 0.0321 (0.000) & 0.0232 (0.000) & 0.0402 (0.000) & 0.0269 (0.000) \\
\textbf{age\_group}              & 0.0006 (0.000) & 0.0005 (0.220) & 0.0075 (0.000) & 0.0005 (0.299) & 0.0013 (0.000) \\
\textbf{gender}                  & 0.0001 (0.133) & 0.0011 (0.000) & 0.0005 (0.005) & 0.0003 (0.078) & 0.0001 (0.643) \\
\textbf{education}               & 0.0117 (0.000) & 0.0050 (0.000) & 0.0005 (0.003) & 0.0012 (0.000) & 0.0028 (0.000) \\
\textbf{city\_size}              & 0.0007 (0.000) & 0.0011 (0.001) & 0.0020 (0.000) & 0.0007 (0.065) & 0.0019 (0.000) \\
\textbf{econ\_now}               & 0.0004 (0.000) & 0.0004 (0.173) & 0.0011 (0.000) & 0.0011 (0.000) & 0.0010 (0.000) \\
\textbf{econ\_future}            & 0.0027 (0.000) & 0.0040 (0.000) & 0.0009 (0.000) & 0.0013 (0.000) & 0.0023 (0.000) \\
\textbf{econ\_personal\_future}  & 0.0008 (0.000) & 0.0027 (0.000) & 0.0014 (0.000) & 0.0008 (0.007) & 0.0018 (0.000) \\
\textbf{trust\_people}           & 0.0001 (0.004) & 0.0002 (0.185) & 0.0000 (0.791) & 0.0012 (0.000) & 0.0002 (0.207) \\
\textbf{democracy\_pref}         & 0.0003 (0.000) & 0.0033 (0.000) & 0.0009 (0.000) & 0.0019 (0.000) & 0.0009 (0.001) \\
\textbf{demo\_satisfaction}      & 0.0008 (0.000) & 0.0005 (0.042) & 0.0012 (0.000) & 0.0010 (0.001) & 0.0017 (0.000) \\
\textbf{trust\_parliament}       & 0.0005 (0.000) & 0.0008 (0.005) & 0.0001 (0.722) & 0.0009 (0.003) & 0.0008 (0.003) \\
\textbf{trust\_gov}              & 0.0010 (0.000) & 0.0003 (0.223) & 0.0004 (0.059) & 0.0010 (0.001) & 0.0012 (0.000) \\
\textbf{trust\_judiciary}        & 0.0006 (0.000) & 0.0007 (0.011) & 0.0002 (0.363) & 0.0007 (0.010) & 0.0011 (0.000) \\
\textbf{ideology\_group}         & 0.0009 (0.000) & 0.0035 (0.000) & 0.0011 (0.000) & 0.0005 (0.057) & 0.0021 (0.000) \\
\bottomrule
\end{tabular}
\caption{Mutual information (MI) between \textbf{fear\_ai} and each feature, with p-values from Fisher’s exact test (in parentheses). Age was discretized into quartiles prior to analysis. All MI values are rounded to 4 decimal digits, and p-values to 3.}
\label{tab:mi}
\end{table*}

 \subsection{Associations Between Predictors and Fear of AI}

Table~\ref{tab:mi} reports the association between \textbf{fear\_ai} and each predictor, as measured by mutual information (MI) and tested via Fisher's exact test. The features most strongly associated with concern about AI-driven job loss are \textbf{education} and \textbf{country}, followed by \textbf{econ\_future} and \textbf{demo\_satisfaction}. These variables exhibit consistently higher MI values and highly significant p-values across all years.

Discretized \textbf{age\_group} shows modest association with \textbf{fear\_ai}, particularly in 2018, but is weaker in other years. In contrast, variables like \textbf{gender} and \textbf{trust\_people} exhibit very low MI values despite occasionally significant p-values, suggesting that these variables contribute little explanatory power, despite some statistical significance. Overall, the results highlight that structural (e.g., education, economic outlook) and contextual (e.g., country, political ideology) factors are more informative than basic demographics or general attitudes.

\begin{figure*}
  \centering
  \begin{subfigure}{0.48\textwidth}
    \includegraphics{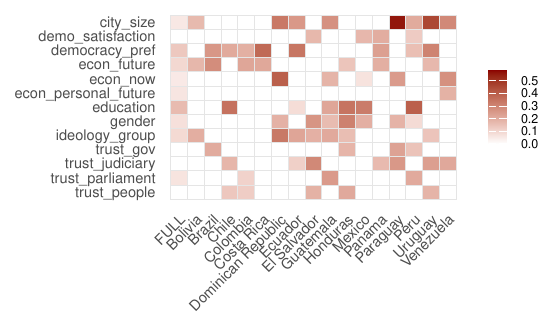}
    \caption{2017}
  \end{subfigure}
  \hfill
  \begin{subfigure}{0.48\textwidth}
    \includegraphics{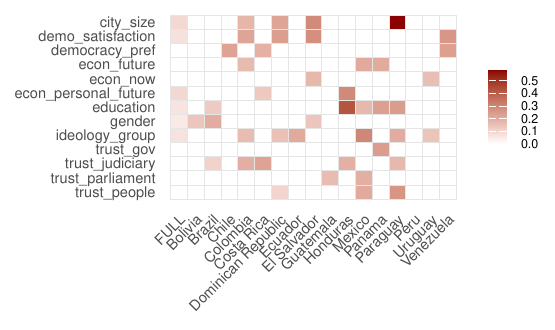}
    \caption{2018}
  \end{subfigure}
  
  \vspace{0.5cm}
  
  \begin{subfigure}{0.48\textwidth}
    \includegraphics{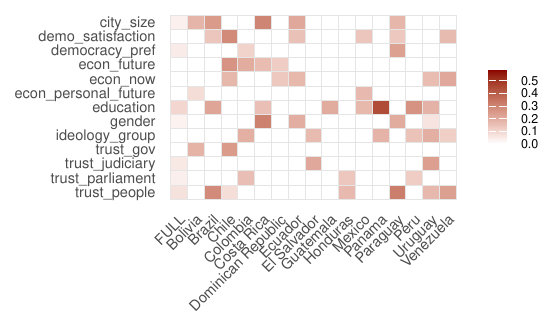}
    \caption{2020}
  \end{subfigure}
  \hfill
  \begin{subfigure}{0.48\textwidth}
    \includegraphics{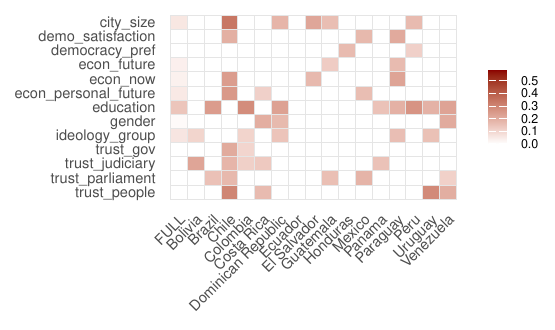}
    \caption{2023}
  \end{subfigure}
  
  \caption{Variable inclusion and strength across countries for each year. White indicates the variable was excluded; darker shades represent higher average absolute coefficients.}
  \label{fig:heatmap_grid}
\end{figure*}

\subsection{Analysis of Covariate Effects Across Years}

Figure~\ref{fig:heatmap_grid} presents heatmaps summarizing the average absolute effect sizes of selected predictors across countries and years, based on cross-validated ordinal regression models. Darker shading corresponds to stronger covariate effects. There is a clear temporal pattern: in 2017, many covariates exhibited non-negligible effects across multiple countries, suggesting a broader range of predictors associated with AI-related job loss fear. In later years (2018, 2020, and 2023), the heatmaps are sparser with fewer variables consistently selected and smaller average effect sizes overall. This suggests that over time, attitudes toward AI job displacement became structured around a smaller set of key predictors.

Across all years, \textbf{education} emerges as the most robust and consistently strong predictor, appearing with relatively high intensity in nearly all countries and years. \textbf{Ideology\_group} also maintains substantial importance across time, indicating that political orientation plays a durable role in shaping views about AI and automation.

\textbf{Gender} shows moderate importance in 2017 but diminishes almost entirely in later years, suggesting that gender differences in AI fear diminished over time. Conversely, \textbf{city\_size} displays relatively stable importance across all four waves, particularly in countries such as Uruguay, Costa Rica, and Mexico, where urban-rural divides appear to influence perceptions of technological disruption.

Interestingly, \textbf{democracy\_pref} has strong effects in 2017 but almost disappears as a predictor in subsequent years. This pattern may reflect changing political narratives around technology, or the relative decline in the salience of democratic preferences for structuring attitudes toward AI after the initial surge of attention in 2017–2018.

Comparing across countries, certain nations exhibit distinct profiles. For instance, Chile consistently shows strong effects for trust-related variables, including \textbf{trust\_gov} and \textbf{trust\_judiciary}, suggesting that institutional confidence plays a particularly prominent role in Chilean attitudes toward automation. Colombia, on the other hand, presents a markedly different profile, with \textbf{ideology\_group} dominating and \textbf{education} playing a secondary role, contrary to the broader regional trend.

Overall, the heatmaps show structural stability, such as the persistent influence of education and ideology, and cross-country differences in how social, economic, and political factors shape public fear of AI-driven job displacement.

\subsection{Evolution of Key Predictors Over Time}
\begin{figure*}
    \centering
    \includegraphics{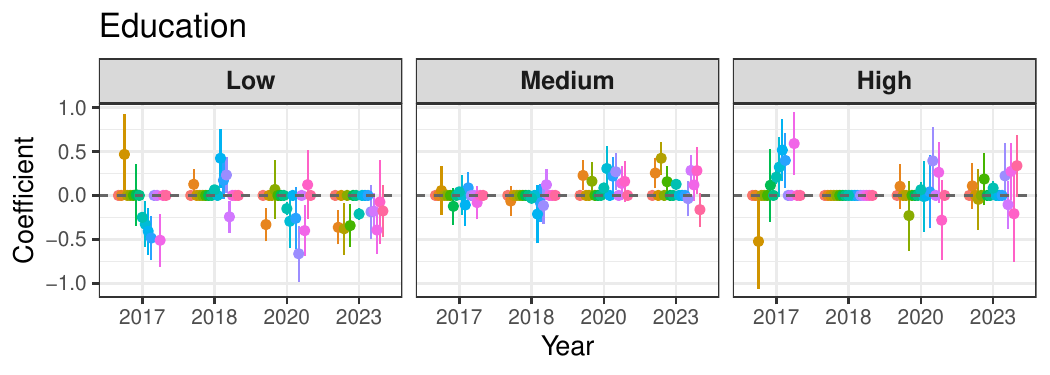}
    \includegraphics{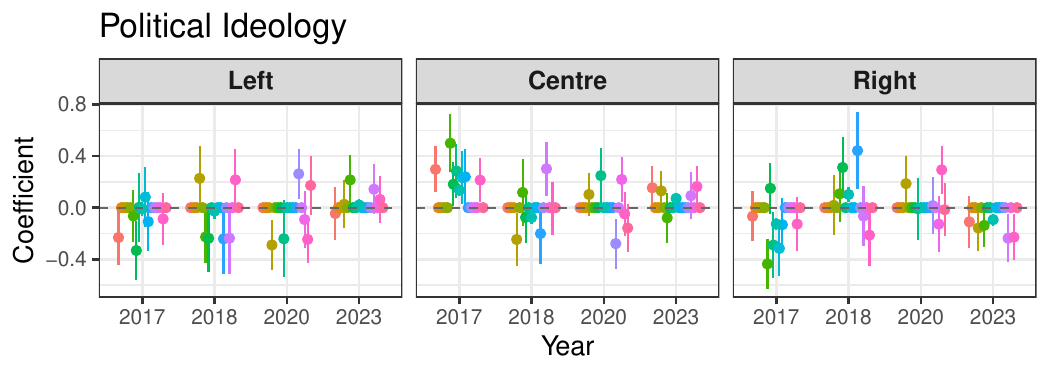}
    \includegraphics{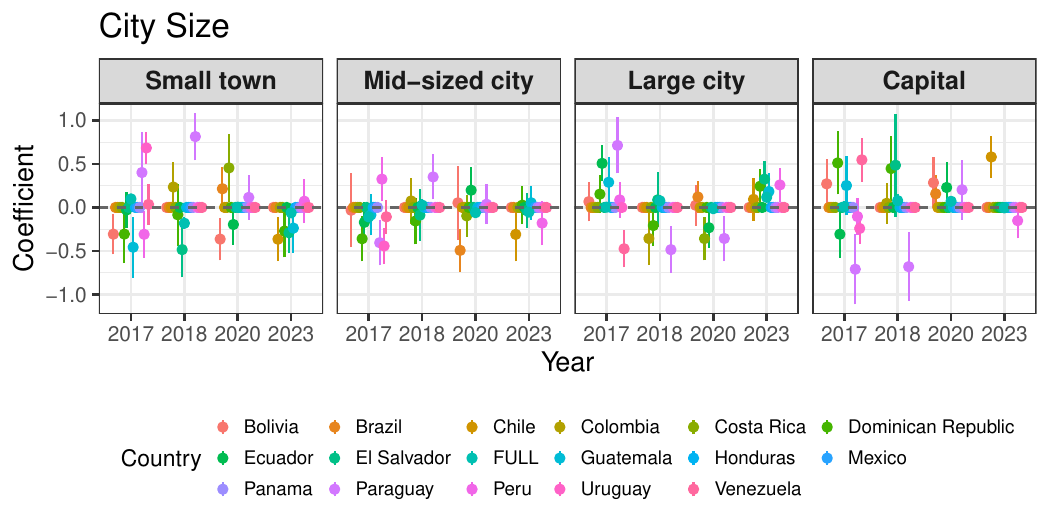}
    \caption{Evolution of estimated coefficients for \textbf{education} (top), \textbf{political ideology} (middle), and \textbf{city size} (bottom) across countries and years. 
Error bars indicate 95\% confidence intervals.
}
    \label{fig:feature_city_size}
\end{figure*}

Figure~\ref{fig:feature_city_size} displays the evolution of estimated coefficients for three key predictors, \textbf{education} (top), \textbf{political ideology} (middle), and \textbf{city size} (bottom), across countries and survey years. Each panel shows country-level estimates for each level of the predictor, with the pooled ``FULL'' model included for comparison. Because of the model structure and the use of sum-to-zero contrasts, negative coefficient estimates indicate greater fear of AI-driven job loss (i.e., a higher probability of agreeing or strongly agreeing that AI will lead to job displacement).

For \textbf{education}, individuals with lower attainment generally exhibit more negative coefficients, indicating a stronger association with fear of AI-driven job loss. Medium and high education levels show near-zero or slightly positive coefficients, consistent with a buffering effect of education.

For \textbf{political ideology}, coefficients are mostly close to zero, though the \textit{Centre} category consistently shows slightly more positive values than Left or Right, suggesting that centrist respondents tend to express less concern about AI-driven job displacement. That said, the direction of coefficients for Left and Right varies by year: while left-leaning individuals show greater concern in earlier waves (e.g., 2017 and 2018), this pattern is less clear or even reversed by 2023.

For \textbf{city size}, respondents in large cities and capitals tend to show more positive coefficients, especially in 2023, indicating less fear of job loss in urban areas. In contrast, those in mid-sized cities and small towns typically show coefficients closer to zero or negative, pointing to a modest but persistent urban–rural divide.

Full-model estimates (available in the extended version) confirm these patterns, with education and ideology showing some of the largest coefficients across years and supporting their role as structural drivers of concern over AI-induced job loss.

\subsection{The Role of Age in Shaping Fear of AI}

Unlike other predictors, \textbf{age} is treated as a continuous variable in the ordinal regression models. It is included in approximately 32\% of fitted models across countries and years, and when selected, the estimated coefficients are predominantly negative, suggesting that younger individuals tend to express greater concern about AI-driven job loss than older individuals. This negative association is most evident in 2017 and 2023, particularly in Bolivia, Costa Rica, El Salvador, and Peru (see the extended version).

The overall effect of age, however, is modest and varies across countries and years. In 2020, for instance, positive coefficients emerge in Mexico, Paraguay, and Guatemala, indicating that in some contexts, older individuals express greater concern. These findings suggest that generational divides exist but are less consistent than those observed for education, political ideology, or city size, and likely reflect context-dependent patterns shaped by national or temporal factors.

\begin{table*}[ht]
\small
\centering
\setlength{\tabcolsep}{1mm}
\begin{tabular}{lcccccccccccccccc}
\toprule
\textbf{Fear of AI} 
& \multicolumn{4}{c}{\textbf{2017}} 
& \multicolumn{4}{c}{\textbf{2018}} 
& \multicolumn{4}{c}{\textbf{2020}} 
& \multicolumn{4}{c}{\textbf{2023}} \\
\cmidrule(lr){2-5} \cmidrule(lr){6-9} \cmidrule(lr){10-13} \cmidrule(lr){14-17}
 & C1 & C2 & C3 & C4 & C1 & C2 & C3 & C4 & C1 & C2 & C3 & C4 & C1 & C2 & C3 & C4 \\
\midrule
Strongly disagree & 3\% & 5\% & 4\% & 3\% & 0\% & 0\% & 0\% & 0\% & 12\% & 14\% & 13\% & 12\% & 13\% & 15\% & 12\% & 14\% \\
Disagree          & 21\% & 21\% & 21\% & 18\% & 18\% & 15\% & 17\% & 16\% & 45\% & 37\% & 42\% & 41\% & 43\% & 37\% & 40\% & 39\% \\
Agree             & 48\% & 45\% & 50\% & 49\% & 69\% & 68\% & 70\% & 70\% & 31\% & 35\% & 34\% & 35\% & 30\% & 31\% & 34\% & 32\% \\
Strongly agree    & 28\% & 29\% & 26\% & 30\% & 14\% & 17\% & 14\% & 14\% & 11\% & 13\% & 11\% & 12\% & 15\% & 17\% & 15\% & 16\% \\
\bottomrule
\end{tabular}
\caption{Distribution of \textbf{fear\_ai} responses (in \%) across latent classes and survey years. }
\label{tab:fearai_by_class_year}
\end{table*}

\begin{table}
\small
\centering
\setlength{\tabcolsep}{1mm}
\begin{tabular}{lcccc}
\toprule
\textbf{Class Pair} & \textbf{2017} & \textbf{2018} & \textbf{2020} & \textbf{2023} \\
\midrule
C1 vs C2 & 0.009 & 0.005 & 0.001 & 0.002 \\
C1 vs C3 & 0.192 & 0.506 & 0.114     & 0.005 \\
C1 vs C4 & 0.032 & 0.360 & 0.011     & 0.116 \\
C2 vs C3 & 0.005 & 0.005 & 0.003     & 0.002 \\
C2 vs C4 & 0.009 & 0.016 & 0.016     & 0.355 \\
C3 vs C4 & 0.005 & 0.881 & 0.439     & 0.145 \\
\bottomrule
\end{tabular}
\caption{Adjusted $p$-values (with Benjamini--Hochberg procedure) from pairwise chi-squared tests comparing \textbf{fear\_ai} response distributions between latent classes, for each year.}

\label{tab:lca_pairwise_summary}
\end{table}

\subsection{Latent Class Analysis of Respondent Profiles}

To identify distinct groups of respondents based on their demographic characteristics, economic perceptions, political attitudes, and trust in institutions, we perform a Latent Class Analysis (LCA) using all predictors except \textbf{country} and \textbf{fear\_ai}. Although the BIC decreases monotonically with additional classes, likely due to the large sample size, models with more than four classes show diminishing separation and interpretability. We therefore retain a four-class solution as a compromise between BIC and entropy, balancing model fit and classification certainty (see the extended version for details).

Each class reflects a distinct combination of respondent characteristics, summarized below. Interpretations are based on the estimated item-response probabilities \( \mathbb{P}(X_j = x_j \mid C = k) \), which represent the likelihood that individuals in latent class \( k \) choose a particular category of variable \( X_j \). These probabilities inform the substantive meaning of each profile and are available in the extended version.

\begin{itemize}
    \item \textbf{C1: Optimistic Institutionalists}. Respondents in this group tend to report high trust in government, parliament, and the judiciary, as well as relatively high satisfaction with democracy. They display strong economic optimism, particularly regarding future personal and national conditions, and are evenly distributed across age groups, genders, and education levels.

    \item \textbf{C2: Disillusioned Pessimists}. This group is characterized by overwhelmingly negative views of both the current and future economy, very low institutional trust, and high dissatisfaction with democracy. Respondents are somewhat older and more likely to live in small or mid-sized towns. They also report the highest generalized distrust in other people.

    \item \textbf{C3: Moderate Progressives}. These respondents tend to be younger, with above-average levels of education and moderate trust in institutions. They report moderate satisfaction with democracy and relatively good expectations for their personal economic future. Their ideological self-placement is more often centrist.

    \item \textbf{C4: Distrustful Urbanites}. This group includes respondents concentrated in large cities or capitals. They have low institutional trust, particularly in government and the judiciary, and a mixed economic outlook, more optimistic about their personal than national future. Their education levels are above average, and they endorse democracy at relatively high rates despite reporting dissatisfaction with its performance.

\end{itemize}

We begin by testing whether the distribution of \textbf{fear\_ai} responses differs significantly across the four latent classes within each survey year. Global chi-squared tests confirm strong evidence of association in all years (2017–2023), with $p$-values below 0.01 in every case. These results indicate that, although overall fear of AI is widespread, its expression varies systematically across respondent profiles.

To better understand these differences, Table~\ref{tab:fearai_by_class_year} reports the percentage of responses in each fear category, by class and year. The differences in proportions appear modest, often just a few percentage points, but remain statistically significant in many comparisons, as shown in the pairwise tests reported in Table~\ref{tab:lca_pairwise_summary}. For instance, in 2023, responses of “Strongly agree” range from 15\% (C1) to 17\% (C2), yet several class pairs show $p$-values below 0.01. 

These subtle but robust differences align closely with the class profiles. \textbf{C2 (Disillusioned Pessimists)} consistently report the highest levels of concern about AI-induced job loss, reflecting their economic pessimism and institutional distrust. \textbf{C1 (Optimistic Institutionalists)} express the lowest concern, particularly in 2020 and 2023. \textbf{C3 (Moderate Progressives)} and \textbf{C4 (Distrustful Urbanites)} fall in between, but differ in political trust and urban exposure, with the latter group showing greater concern in more recent years. 

\begin{figure*}
\centering
\includegraphics{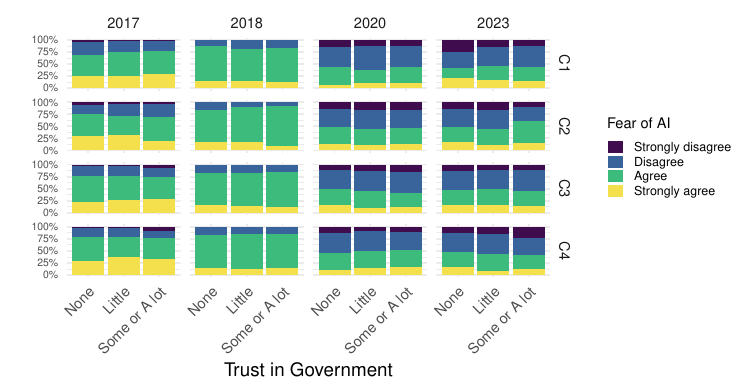}
\caption{Distribution of \textbf{fear\_ai} responses across levels of trust in government, stratified by latent class and survey year. }
\label{fig:fearai_trustgov}
\end{figure*}

To assess whether latent class membership captures attitudinal patterns that go beyond observed covariates, we examine how fear of AI-driven job loss varies with institutional trust across classes. Trust in government serves as a representative example, given its prominence in prior literature on AI perceptions. As shown in Figure~\ref{fig:fearai_trustgov}, variation in fear levels is primarily explained by class membership rather than differences in trust alone. \textbf{C1 (Optimistic Institutionalists)} consistently reports the lowest concern, with a higher share of “Disagree” or “Strongly disagree” responses across all levels of trust. \textbf{C2 (Disillusioned Pessimists)}, by contrast, shows the highest fear regardless of reported trust. \textbf{C3 (Moderate Progressives)} and \textbf{C4 (Distrustful Urbanites)} occupy intermediate positions but differ in both baseline fear and sensitivity to trust. These patterns suggest that the latent classes capture stable attitudinal configurations not reducible to any single predictor.

\section{Discussion}

This study offers a comprehensive, multi-country analysis of how fear of AI-induced job loss has evolved in Latin America. Our findings reveal both widespread concern and notable heterogeneity in how these fears are structured across time, geography, and social groups.

Consistent with prior research in high-income countries, we find that education is the most consistent predictor of AI-related fear across time and countries \citep{bergdahl2023selfdetermination, schiavo2024comprehension}. Individuals with lower education levels are more likely to express concern about losing their job to AI and automation, while those with higher education levels report lower fear \citep{yarovenko2024future}. This pattern is evident in both the cross-country heatmaps and the full models, where education consistently exhibits some of the largest coefficients across years.

Political ideology plays a nuanced role in shaping attitudes toward AI. Across most years, respondents identifying with the political centre tend to express slightly less concern about AI-driven job loss compared to those on the ideological left or right. While earlier waves (e.g., 2017 and 2018) show somewhat greater fear among left-leaning individuals, this pattern becomes less consistent over time, with right-leaning respondents exhibiting greater concern in some countries by 2023. These findings partially mirror results from European studies such as \citet{gerlich2023perceptions}, which highlight ideological divides in perceptions of automation, though in our data the gradient is modest and appears to shift over time rather than follow a fixed partisan pattern.

By contrast, the role of economic perceptions is more mixed. Although both national and personal economic outlooks are occasionally selected as predictors in the yearly models, their coefficients tend to be smaller and less consistent over time. This diverges from some findings in high-income contexts, where economic pessimism is a stronger driver of automation-related anxiety \citep{frey2017future}. Similarly, while institutional trust, particularly in government, parliament, and the judiciary, is included in some country-specific models, it plays a more secondary role in the pooled yearly estimates. However, the latent class results indicate that trust remains relevant for certain groups: \textbf{C2 (Disillusioned Pessimists)}, for instance, combine high concern about AI with persistent institutional distrust.

Latent Class Analysis further reveals that societal perceptions of AI cluster into distinct attitudinal profiles that extend beyond individual predictors. While all classes exhibit elevated concern in 2018, subsequent waves show more differentiated patterns. \textbf{C2 (Disillusioned Pessimists)} consistently express the highest fear, whereas \textbf{C1 (Optimistic Institutionalists)} show the lowest, particularly in 2020 and 2023. \textbf{C3 (Moderate Progressives)} and \textbf{C4 (Distrustful Urbanites)} fall in between, though they differ in their levels of institutional trust and urban exposure. These patterns suggest that fear of automation is shaped not only by observable traits but also by more stable attitudinal configurations that persist across survey years.

Overall, our findings show that fear of automation in Latin America cannot be reduced to a single explanation or confined to particular demographic categories. Rather, it reflects a dynamic interaction between education, political beliefs, economic expectations, and institutional trust. As AI technologies continue to advance, these fears may intensify or shift depending on how they intersect with broader trends in inequality, labor markets, and political identity.

\section{Limitations}

Several limitations should be considered when interpreting our findings. First, the data are observational and cross-sectional, which precludes causal inference. While we identify strong associations between predictors and fear of AI, we cannot determine the directionality of these relationships. Second, although we harmonized the outcome variable across survey waves, the wording of the question on AI-related job loss changed slightly over time. In particular, the 2018 item referenced the risk to both the respondent and their family, which likely contributed to the unusually high agreement observed in that year. Third, the statistical models rely on self-reported perceptions, which may be subject to reporting biases, particularly in politically sensitive environments. Finally, while our full models provide generalizable trends across the region, they may obscure meaningful variation at the country level. For this reason, we included country-specific heatmaps and latent class profiles to highlight diversity in attitudes. Future work could incorporate panel data or event-based designs to capture shifts in response to economic or political developments.

\section{Conclusions}

This study contributes to the literature on public perceptions of AI by analyzing fear of job loss across four waves of the Latinobarómetro survey in 16 Latin American countries. Our findings highlight that concern about AI-driven automation is widespread, particularly among individuals with lower education levels and those identifying with the political left. Temporal and cross-country variation suggests that fear is shaped not only by individual characteristics but also by ideological context. Latent class analysis reveals four distinct respondent profiles, reflecting different combinations of demographic, economic, and political traits.

These findings have practical relevance for governments and organizations seeking to implement AI. Addressing public concern will require not only technical safeguards but also broader efforts to build institutional trust, reduce structural inequalities, and promote inclusive communication around the risks and benefits of automation. Policies aimed at upskilling the workforce, increasing transparency, and engaging citizens in AI governance may help mitigate fear and foster equitable technological adoption.  These results underscore the need to address AI anxiety not as a technological issue, but as a social and political challenge shaped by education, trust, and inequality.

\section{Acknowledgments}

This paper was partially funded by PID2023-153222OB-I00 granted by MCIU / AEI / 10.13039/501100011033 / FEDER, UE.

\bibliography{aaai25}

\appendix

\section{Additional Results}
\label{sec:lca_appendix}
\begin{table*}[ht]
\centering
\begin{tabular}{llrrrr}
\toprule
Feature & Level & 2017 & 2018 & 2020 & 2023 \\
\midrule
\textbf{age} & Age (continuous) &  & -0.003 &  & -0.002 \\
\textbf{education} & Low & -0.249 & 0.065 & -0.150 & -0.211 \\
\textbf{education} & Medium & 0.041 &  & 0.085 & 0.127 \\
\textbf{education} & High &0.208 && 0.065 & 0.084\\ 
\textbf{gender} & Female & -0.074 & -0.051 & 0.028 & 0.012 \\
\textbf{gender} & Male & 0.074 & 0.051 & -0.028 & -0.012 \\
\textbf{city\_size} & Small town & 0.098 & -0.182 & 0.003 & -0.061 \\
\textbf{city\_size} & Mid-sized city & -0.109 & 0.027 & -0.060 & -0.052 \\
\textbf{city\_size} & Large city & 0.003 & 0.076 & -0.014 & 0.117 \\
\textbf{city\_size} & Capital& 0.008 & 0.079 & 0.071 & -0.004 \\
\textbf{econ\_future} & Worse & -0.014 &  &  & 0.014 \\
\textbf{econ\_future} & Same & 0.143 &  &  & 0.042 \\
\textbf{econ\_future} & Better & -0.129 &  &  & -0.056 \\
\textbf{econ\_personal\_future} & Worse & -0.073 & 0.127 &  & 0.080 \\
\textbf{econ\_personal\_future} & Same & 0.093 & -0.140 &  & -0.036 \\
\textbf{econ\_personal\_future} & Better & -0.020 & 0.013 &  & -0.044 \\
\textbf{econ\_now} & Bad & 0.015 &  &  & 0.027 \\
\textbf{econ\_now} & Average & -0.078 &  &  & -0.051 \\
\textbf{econ\_now} & Better & 0.063 &  &  & 0.024 \\
\textbf{ideology\_group} & Left & -0.010 & -0.027 &  & 0.020 \\
\textbf{ideology\_group} & Centre & 0.135 & -0.076 &  & 0.072 \\
\textbf{ideology\_group} & Right & -0.125 & 0.103 &  & -0.092 \\
\textbf{democracy\_pref} & Doesn't matter & -0.200 &  & 0.040 &  \\
\textbf{democracy\_pref} & Authoritarian OK & 0.076 &  & 0.033 &  \\
\textbf{democracy\_pref} & Democracy best & 0.124 &  & -0.073 &  \\
\textbf{demo\_satisfaction} & Not at all satisfied &  & 0.105 &  &  \\
\textbf{demo\_satisfaction} & Not very satisfied &  & -0.007 &  &  \\
\textbf{demo\_satisfaction} & Satisfied &  & -0.098 &  &  \\
\textbf{trust\_parliament} & None & 0.098 &  & -0.006 &  \\
\textbf{trust\_parliament} & Little & -0.063 &  & -0.052 &  \\
\textbf{trust\_parliament} & Some or A lot & -0.035 &  & 0.058 &  \\
\textbf{trust\_people} & Cannot be too careful &  &  & 0.071 &  \\
\textbf{trust\_people} & Trust most &  &  & -0.071 &  \\
\textbf{trust\_judiciary} & None &  &  & 0.052 &  \\
\textbf{trust\_judiciary} & Little &  &  & 0.028 &  \\
\textbf{trust\_judiciary} & Some or A lot &  &  & -0.080 &  \\
\bottomrule
\end{tabular}
\caption{Estimated coefficients from ``FULL" models by predictor level and year.}
\label{tab:coef_by_year}
\end{table*}

\begin{table*}
\centering
\begin{tabular}{lcccc}
\toprule
\textbf{Country} & \textbf{2017} & \textbf{2018} & \textbf{2020} & \textbf{2023} \\
\midrule
Bolivia            & -0.0072 &         & -0.0089 & -0.0140 \\
Brazil             &         & -0.0108 &         &         \\
Chile              &         &         & -0.0060 &         \\
Colombia           &         &         & -0.0138 &         \\
Costa Rica         & -0.0151 &         & -0.0067 &         \\
Dominican Republic &         &         &         &         \\
Ecuador            &         &         &         &         \\
El Salvador        & -0.0116 &         &         & -0.0103 \\
FULL               &         & -0.0031 &         & -0.0023 \\
Guatemala          & -0.0138 &         &  0.0148 &         \\
Honduras           &         &         &         &         \\
Mexico             &         &         &  0.0127 &         \\
Panama             &         &         &         & -0.0095 \\
Paraguay           &         &         &  0.0126 &         \\
Peru               & -0.0121 &         &         & -0.0103 \\
Uruguay            &         &  0.0229 &         &         \\
Venezuela          &         & -0.0055 &         &         \\
\bottomrule
\end{tabular}
\caption{Estimated coefficients for \textbf{age} in country-specific and pooled (FULL) models across survey years. Blank cells indicate that \textbf{age} was not included in the model for that country-year.}
\label{tab:age_coefficients}
\end{table*}

\begin{table*}
\centering
\begin{tabular}{ccc}
\toprule
\textbf{Number of Classes} & \textbf{BIC} & \textbf{Entropy} \\
\midrule
1  & $1.390 \times 10^6$ & --    \\
2  & $1.332 \times 10^6$ & 0.767 \\
3  & $1.316 \times 10^6$ & 0.753 \\
4  & $1.309 \times 10^6$ & 0.734 \\
5  & $1.305 \times 10^6$ & 0.714 \\
6  & $1.302 \times 10^6$ & 0.707 \\
7  & $1.300 \times 10^6$ & 0.707 \\
8  & $1.299 \times 10^6$ & 0.689 \\
9  & $1.298 \times 10^6$ & 0.701 \\
10 & $1.296 \times 10^6$ & 0.701 \\
\bottomrule
\end{tabular}
\caption{Bayesian Information Criterion (BIC) and entropy for latent class models with 1 to 10 classes. BIC is reported in scientific notation to one decimal digit. Entropy measures classification certainty, with higher values indicating better separation.}
\label{tab:lca_bic_entropy}
\end{table*}

\begin{table*}
\centering
\begin{tabular}{lcccc}
\toprule
\textbf{Education} & \textbf{Class 1} & \textbf{Class 2} & \textbf{Class 3} & \textbf{Class 4} \\
\midrule
Low    & 0.274 & 0.316 & 0.289 & 0.297 \\
Medium & 0.506 & 0.495 & 0.517 & 0.509 \\
High   & 0.220 & 0.189 & 0.194 & 0.194 \\
\bottomrule
\end{tabular}
\caption{Item-response probabilities for \textbf{education} across latent classes.}
\label{tab:lca_education}
\end{table*}

\begin{table*}
\centering
\begin{tabular}{lcccc}
\toprule
\textbf{Age Group} & \textbf{Class 1} & \textbf{Class 2} & \textbf{Class 3} & \textbf{Class 4} \\
\midrule
Q1 (Youngest) & 0.241 & 0.147 & 0.302 & 0.263 \\
Q2            & 0.222 & 0.235 & 0.258 & 0.283 \\
Q3            & 0.238 & 0.295 & 0.232 & 0.255 \\
Q4 (Oldest)   & 0.299 & 0.324 & 0.208 & 0.199 \\
\bottomrule
\end{tabular}
\caption{Item-response probabilities for \textbf{age\_group} across latent classes.}
\label{tab:lca_agegroup}
\end{table*}

\begin{table*}
\centering
\begin{tabular}{lcccc}
\toprule
\textbf{Gender} & \textbf{Class 1} & \textbf{Class 2} & \textbf{Class 3} & \textbf{Class 4} \\
\midrule
Female & 0.463 & 0.520 & 0.504 & 0.481 \\
Male   & 0.537 & 0.480 & 0.496 & 0.519 \\
\bottomrule
\end{tabular}
\caption{Item-response probabilities for \textbf{gender} across latent classes.}
\label{tab:lca_gender}
\end{table*}

\begin{table*}
\centering
\begin{tabular}{lcccc}
\toprule
\textbf{City Size} & \textbf{Class 1} & \textbf{Class 2} & \textbf{Class 3} & \textbf{Class 4} \\
\midrule
Small Town     & 0.252 & 0.176 & 0.220 & 0.172 \\
Mid-sized City & 0.283 & 0.284 & 0.267 & 0.254 \\
Large City     & 0.274 & 0.339 & 0.332 & 0.357 \\
Capital        & 0.191 & 0.201 & 0.181 & 0.217 \\
\bottomrule
\end{tabular}
\caption{Item-response probabilities for \textbf{city\_size} across latent classes.}
\label{tab:lca_citysize}
\end{table*}

\begin{table*}
\centering
\begin{tabular}{lcccc}
\toprule
\textbf{Current Economic Perception} & \textbf{Class 1} & \textbf{Class 2} & \textbf{Class 3} & \textbf{Class 4} \\
\midrule
Bad     & 0.127 & 0.825 & 0.337 & 0.463 \\
Average & 0.557 & 0.158 & 0.571 & 0.457 \\
Good    & 0.317 & 0.017 & 0.092 & 0.079 \\
\bottomrule
\end{tabular}
\caption{Item-response probabilities for \textbf{econ\_now} across latent classes.}
\label{tab:lca_econnow}
\end{table*}

\begin{table*}
\centering
\begin{tabular}{lcccc}
\toprule
\textbf{Future Economic Outlook} & \textbf{Class 1} & \textbf{Class 2} & \textbf{Class 3} & \textbf{Class 4} \\
\midrule
Worse  & 0.092 & 0.845 & 0.246 & 0.156 \\
Same   & 0.252 & 0.130 & 0.386 & 0.446 \\
Better & 0.657 & 0.025 & 0.368 & 0.398 \\
\bottomrule
\end{tabular}
\caption{Item-response probabilities for \textbf{econ\_future} across latent classes.}
\label{tab:lca_econfuture}
\end{table*}

\begin{table*}
\centering
\begin{tabular}{lcccc}
\toprule
\textbf{Personal Economic Outlook} & \textbf{Class 1} & \textbf{Class 2} & \textbf{Class 3} & \textbf{Class 4} \\
\midrule
Worse  & 0.046 & 0.567 & 0.122 & 0.025 \\
Same   & 0.263 & 0.278 & 0.360 & 0.368 \\
Better & 0.691 & 0.155 & 0.517 & 0.607 \\
\bottomrule
\end{tabular}
\caption{Item-response probabilities for \textbf{econ\_personal\_future} across latent classes.}
\label{tab:lca_econpersonal}
\end{table*}

\begin{table*}
\centering
\begin{tabular}{lcccc}
\toprule
\textbf{Trust in People} & \textbf{Class 1} & \textbf{Class 2} & \textbf{Class 3} & \textbf{Class 4} \\
\midrule
Cannot be too careful     & 0.765 & 0.908 & 0.859 & 0.881 \\
Most people can be trusted & 0.235 & 0.092 & 0.141 & 0.119 \\
\bottomrule
\end{tabular}
\caption{Item-response probabilities for \textbf{trust\_people} across latent classes.}
\label{tab:lca_trustpeople}
\end{table*}

\begin{table*}
\centering
\begin{tabular}{lcccc}
\toprule
\textbf{Trust in Government} & \textbf{Class 1} & \textbf{Class 2} & \textbf{Class 3} & \textbf{Class 4} \\
\midrule
None           & 0.025 & 0.854 & 0.115 & 0.832 \\
Little         & 0.124 & 0.104 & 0.772 & 0.118 \\
Some or A lot  & 0.851 & 0.042 & 0.113 & 0.050 \\
\bottomrule
\end{tabular}
\caption{Item-response probabilities for \textbf{trust\_gov} across latent classes.}
\label{tab:lca_trustgov}
\end{table*}

\begin{table*}
\centering
\begin{tabular}{lcccc}
\toprule
\textbf{Trust in Parliament} & \textbf{Class 1} & \textbf{Class 2} & \textbf{Class 3} & \textbf{Class 4} \\
\midrule
None           & 0.107 & 0.726 & 0.136 & 0.822 \\
Little         & 0.249 & 0.187 & 0.721 & 0.136 \\
Some or A lot  & 0.644 & 0.087 & 0.144 & 0.042 \\
\bottomrule
\end{tabular}
\caption{Item-response probabilities for \textbf{trust\_parliament} across latent classes.}
\label{tab:lca_trustparliament}
\end{table*}

\begin{table*}
\centering
\begin{tabular}{lcccc}
\toprule
\textbf{Trust in Judiciary} & \textbf{Class 1} & \textbf{Class 2} & \textbf{Class 3} & \textbf{Class 4} \\
\midrule
None           & 0.074 & 0.699 & 0.093 & 0.724 \\
Little         & 0.219 & 0.207 & 0.692 & 0.191 \\
Some or A lot  & 0.707 & 0.094 & 0.215 & 0.085 \\
\bottomrule
\end{tabular}
\caption{Item-response probabilities for \textbf{trust\_judiciary} across latent classes.}
\label{tab:lca_trustjudiciary}
\end{table*}

\begin{table*}
\centering
\begin{tabular}{lcccc}
\toprule
\textbf{Democracy Preference} & \textbf{Class 1} & \textbf{Class 2} & \textbf{Class 3} & \textbf{Class 4} \\
\midrule
Does not matter  & 0.169 & 0.347 & 0.302 & 0.353 \\
Authoritarian OK & 0.147 & 0.142 & 0.182 & 0.172 \\
Democracy best   & 0.684 & 0.511 & 0.516 & 0.474 \\
\bottomrule
\end{tabular}
\caption{Item-response probabilities for \textbf{democracy\_pref} across latent classes.}
\label{tab:lca_demopref}
\end{table*}

\begin{table*}
\centering
\begin{tabular}{lcccc}
\toprule
\textbf{Satisfaction with Democracy} & \textbf{Class 1} & \textbf{Class 2} & \textbf{Class 3} & \textbf{Class 4} \\
\midrule
Not at all satisfied & 0.049 & 0.622 & 0.185 & 0.370 \\
Not very satisfied   & 0.295 & 0.318 & 0.590 & 0.480 \\
Satisfied            & 0.656 & 0.060 & 0.225 & 0.150 \\
\bottomrule
\end{tabular}
\caption{Item-response probabilities for \textbf{demo\_satisfaction} across latent classes.}
\label{tab:lca_demosat}
\end{table*}

\begin{table*}
\centering
\begin{tabular}{lcccc}
\toprule
\textbf{Ideology Group} & \textbf{Class 1} & \textbf{Class 2} & \textbf{Class 3} & \textbf{Class 4} \\
\midrule
Left    & 0.239 & 0.316 & 0.229 & 0.266 \\
Centre  & 0.389 & 0.392 & 0.514 & 0.484 \\
Right   & 0.371 & 0.293 & 0.257 & 0.250 \\
\bottomrule
\end{tabular}
\caption{Item-response probabilities for \textbf{ideology\_group} across latent classes.}
\label{tab:lca_ideology}
\end{table*}

\end{document}